\documentclass[showpacs,amsmath,amssymb,floatfix,prl,twocolumn]{revtex4}
\usepackage{graphicx,amsmath}

\begin{document}

%void void void void void 

\title {Hall detection of time-reversal symmetry breaking under AC electric driving.}

\author{ A.D. Chepelianskii and H. Bouchiat }
\affiliation{Univ. Paris-Sud, CNRS, UMR 8502, F-91405, Orsay, France }
% \affiliation{$(b)$ Laboratoire de Photonique et de Nanostructures (LPN)-CNRS, route de Nozay, 91460 Marcoussis, France }

\begin{abstract}
In a four terminal sample microscopic time-reversibility leads to symmetry relations 
between resistance measurements where the role of current and voltage leads are exchanged.
These reciprocity relations are a manifestation of general Onsager-Casimir symmetries in equilibrium systems.
We investigate experimentally the validity of time reversal symmetry in a $GaAs/Ga_{1-x}Al_xAs$ Hall bar 
irradiated by an external AC field, at zero magnetic field. 
For inhomogeneous AC fields we find strong deviations from reciprocity relations 
and show that their origin can be understood from the the billiard model of a Hall junction. 
Under homogeneous irradiation the symmetry is more robust, indicating that time-reversal symmetry is preserved.
\end{abstract}

\maketitle

The Onsager-Casimir relations are a consequence at a macroscopic scale 
of microscopic time-reversal symmetry \cite{Onsager,Casimir}. 
In mesoscopic physics, these relations proved crucial in the understanding 
of magnetotransport properties when 
the reciprocity relation between resistances $R$ and $R^*$ measured in experiments 
which exchange the current and voltage leads: 
$R(H) = R^*(-H)$ was derived \cite{Buttiker} and verified experimentally \cite{webb}.
Since then the possibility to extend the reciprocity relation to out-of equilibrium conductors
has attracted considerable attention. 
In the special case where the conductor has only two contacts the reciprocity relation implies that 
the transport is symmetrical with magnetic field $H$: $R(H) = R(-H)$. 
In the nonlinear transport regime, it was predicted theoretically that two terminal 
transport can be asymmetric with magnetic field \cite{sanchez0,spivak,polianski}, 
providing a signature of time-reversal symmetry breaking.
This fact was later confirmed in several experiments \cite{wei05,leturq06,marcus06,angers06,angers08}, and lead to new 
theoretical proposals for the generalization of reciprocal relations to nonlinear transport \cite{forster,astumian}.
Recently it was proposed that time-reversal symmetry breaking can be analyzed from linear dc-magnetotransport 
of a system coupled to non-equilibrium baths \cite{sanchez1}, however in this model 
the presence of a magnetic field is necessary to reveal the breaking of reciprocity relations. 
Other manifestations of time-reversal symmetry breaking in non-equilibrium conductors 
at zero magnetic fields were predicted including commensurability effects 
in the frequency domain \cite{kravtsov} and generation of stationary orbital magnetism \cite{toulouse}.
However to our knowledge these effects have not yet been observed experimentally.
In this Letter we directly probe experimentally time-reversal symmetry in zero magnetic field by measuring 
deviations from the four terminal reciprocity relations in a Hall geometry. 
We interpret our results by extending a billiard model initially developed by Beenakker {\it et.al} \cite{Beenakker}. 
to describe both dc-magnetic field behavior and the influence of an inhomogeneous time 
dependent potential.  

\begin{figure}
\begin{center}
\includegraphics[clip=true,width=9cm]{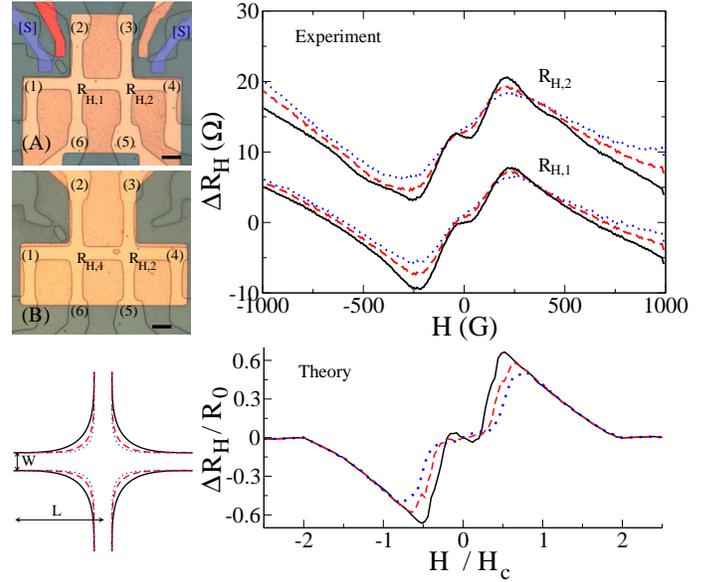}
%\leavevmode
%\epsfxsize=8cm
\caption{(Color online) Left panel, optical micrograph of samples $(A)$ and $(B)$. The black scale line correspond to $10\;{\mu m}$. 
Right panel, on the top we show the experimental behavior of : $\Delta R_H = R_H - \frac{H}{n e}$ as a function of magnetic field for different 
top gate voltages $V_g = 0.1, 0.2, 0.4$ (doted , dashed and continuous line respectively), on the bottom 
theoretical predictions for $\Delta R_H$ for Hall junctions with increasing
curvatures (doted, dashed and continuous line). The specular walls are sketched on the bottom left panel. Temperature was $0.3\;{\rm K}$. 
\label{magneticfield}}
\end{center}
\end{figure}

We have investigated two Hall bars fabricated in a ${\rm GaAs/Ga_{1-x}Al_{x}As}$ two dimensional electron gas (2DEG)
with density  $n_e \simeq 1.2 \times 10^{11}\;{\rm cm^{-2}}$ and mobility $\mu \simeq 1.2 \times 10^2\;{\rm m^2/Vs}$.
The two samples, $(A)$ and $(B)$, were fabricated using wet etching and an aluminum mask.
They have six Au/Ge ohmic contacts to 2DEG  labeled $(1)-(6)$ (see Fig.~1). Both samples are covered by a copper top-gate allowing 
to modulate carrier density. On sample $(A)$ we have also fabricated a local split gate $[S]$ connected to a high 
frequency transmission line; this gate can produce an AC field inhomogeneous on the micron scale.
We define the four terminal resistances $R_{ij,kl}$ as
$R_{ij,kl} = (V_k - V_l) / I_i$ where $V_k$ and $V_l$ are the voltages on the leads $k$ and $l$ 
and $I_i = -I_j$ is the current injected to the source lead $i$. The resistance measured in the configuration where
current and voltage leads are interchanged is noted $R_{ij,kl}^* =  R_{kl,ij}$.
With this notations the reciprocity relation reads $R_{ij,kl}^*(H) = R_{ij,kl}(-H)$ \cite{Buttiker}.
As expected the magnetic field dependence of the Hall resistances $R_{H,1} = R_{14,26}$, $R_{H,2} = R_{14,35}$, 
and of the longitudinal resistance $R_{xx} = R_{14,23}$ exhibits quantum Hall effect plateaux and 
Shubnikov-de Haas oscillations at high magnetic fields $H > 0.5\;{\rm T}$.
The resistances were measured at temperature $T = 0.3 \;{\rm K}$ with an excitation current $I = 1 \;{\mu A}$ modulated at $67\;{\rm Hz}$. 
Voltages were detected with a low noise amplifiers and standard lock-in technique. 
At lower magnetic fields we observe magneto-size peaks, which occur when the size of cyclotron orbits matches 
the width of the 2DEG sample, on both Hall and longitudinal resistance 
for magnetic fields $H \simeq 200\;{\rm G}$.
These observations confirm that our samples are in a ballistic regime. 

The magneto-size peaks on the Hall resistance are analyzed in more detail in Fig.~1 for different top-gate voltages.
To emphasize the magneto-size peaks, we have subtracted the classical Hall resistance $\Delta R_H = R_H - \frac{H}{n_e e}$ 
where $n_e$ is the 2DEG density determined by a linear fit to the Hall resistance at fields above 
the magneto-size peak. An additional cusp appears in $\Delta R_H$  at low magnetic fields 
for higher gate voltages $V_g = 0.4\;{\rm V}$. 
This can be understood from the billiard model of a Hall junction \cite{Beenakker}. 
In this model the Hall junction is treated as a classical billiard with specular walls and four 
contact channels of width $W$ with absorbing boundary conditions at the reservoirs 
distant by $L > W$ 
(possible theoretical geometries are sketched on the left of Fig.~ 1). 
The classical probabilities $P_{i,j}$ of propagating 
from lead $(j)$ to lead $(i)$ are then determined numerically by injecting a large number of classical particles (typically $10^5$) 
at Fermi velocity $v_F$ into lead $j$ and monitoring them until they reach one of the leads $i$. 
The propagation is determined by classical equations of motion in constant field $H$. 
The exit probabilities are then normalized to $\sum_j P_{i,j} = 1$, and the conductance matrix is calculated from: 
\begin{align}
G_{ij} = \frac{1}{R_0} \left[ (1 - P_{ii}) \delta_{ij} + P_{ij} ( 1 - \delta_{ij} ) \right]
\end{align}
Here $R_0 = \frac{h}{2 e^2 N} \simeq \frac{h}{2 e^2} \frac{\pi}{k_F W}$ where $N$ is the channel 
number and $k_F$ the Fermi wavevector. The characteristic magnetic field scale in this model is $H_c = \frac{m v_F}{e W}$, 
where $m$ is the carrier effective mass in 2DEG. 
From the conductance matrix all the four terminal resistances can be calculated including the Hall resistance $R_H$.
On the bottom panel of Fig.~1, we show theoretical magneto-resistances $\Delta R_H$ of  
Hall junctions with different central curvatures. 
For the largest curvature, a cusp appears that is very similar to the behavior observed at higher positive gate voltages. 
When the curvature radius is decreased the cusp disappears, as in the magnetoresistance curves 
at lower gate voltages. These observations suggest that positive gate voltage favor larger curvature radius, 
which is reasonable since higher gate voltages are likely to reduce depletion at the sample boundaries. 
While there is a very good qualitative agreement between the billiard model and our data, 
the agreement is not quantitative. For example for $V_g = 0.4\;{\rm V}$ with 
electron gas density $n_e \approx 2.8 \times 10^{11} {\rm cm}^{-2}$ and estimated channel width $W \simeq 5\;{\rm \mu m}$ 
we find $H_c = \frac{m v_F}{e W}\simeq 170\;{\rm G}$ (we used $v_F = \hbar \sqrt{2 \pi n_e} / m$) .
This leads to a predicted magneto-size peak at 
$H \simeq 100 {\rm G}$ whereas experimentally the peak appears at $H \simeq 250\;{\rm G}$. 

\begin{figure}
\begin{center}
\includegraphics[clip=true,width=9cm]{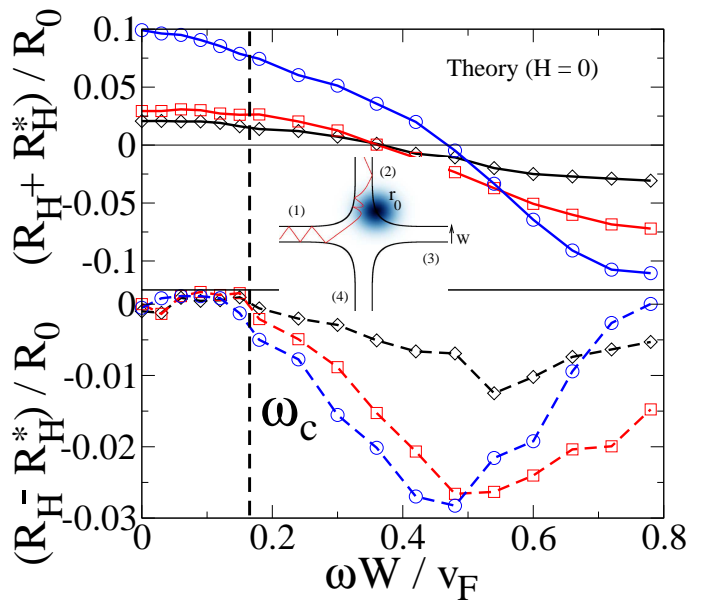}
%\leavevmode
%\epsfxsize=8cm
\caption{(Color online) Calculated dependence of of the Onsager-Casimir symmetric and anti-symmetric components of the 
Hall resistance $R_H$ as a function of the reduced AC-potential frequency $\omega W / v_F$.
Inset shows a ballistic trajectory in presence of an external potential of the form $U(r,t) = U_0 \exp(-k^2 r^2/2) \cos \omega t$ 
centered around $\mathbf{r}_0$ inside the Hall junction with $k W = 1$ (colored circle). 
The anti-symmetric component appears only for $\omega > \omega_c$. 
Symbols represent different potential amplitudes: circles $U_0 = 0.7 \epsilon_F$, squares $U_0 = 0.5 \epsilon_F$ and diamonds $U_0 = 0.35 \epsilon_F$.
}
\label{samplepanddIdVhighV}
\end{center}
\end{figure}

We now address the question of the influence of an external time dependent potential on the Hall resistance,
in zero magnetic field. This problem can be treated theoretically 
if we generalize the billiard model and introduce a localized oscillating potential $U(\mathbf{r}) \cos \omega t$.
As previously the transmission probabilities $P_{i,j}$ are determined by integrating the classical equations of motion
(see typical particle trajectory inset in Fig.~2).
In the static limit $\omega = 0$, an external potential creates a contribution to the 
Hall resistance by deforming the electronic trajectories.
However time reversal symmetry implies that the relation $P_{i,j} = P_{j,i}$ is preserved, and reciprocity relation holds $R_H = R_H^*$. 
Our numerical simulations show however that when the driving frequency $\omega$ is increased the probabilities 
$P_{i,j}$ and $P_{j,i}$ are no longer equal. This causes a difference between $R_H$ and $R_H^*$ even 
at zero magnetic field. For frequencies larger than a certain threshold $\omega_c$ the difference $R_H - R_H^*$ 
becomes of the order of the symmetric contribution  $R_H + R_H^*$. We compare the amplitude of these two  
components as a function of frequency on  Fig.~2. Our simulation shows that 
this frequency is nearly independent on the amplitude of the external potential. 
We note that trajectories that are absorbed in the reservoirs after a single scattering on $U(\mathbf{r})$ 
do not break time reversal symmetry. Indeed in this case it is possible to choose the phase of the external field in a way that 
the time reversed trajectory is also solution of the equations of motion.
As a result the difference between $P_{i,j}$ and $P_{j,i}$ must stem from trajectories that scatter several times on the potential $U(\mathbf{r})$
centered around $\mathbf{r}_0$. 
This allows us to associate the frequency $\omega_c$ with 
the average return time to $\mathbf{r}_0$. 
Interestingly we find in the simulations that the frequency $\omega_c$ is several times smaller than the 
characteristic frequency associated with the size of the channels $v_F / W$.
This points to the role 
of long trajectories with many reflections on the edges of the sample 
with typical length $L_c = \frac{v_F}{\omega_c}$ which can be much larger than $W$, 
of the order of the distance between reservoirs $L$. 

\begin{figure}
\begin{center} 
\includegraphics[clip=true,width=9cm]{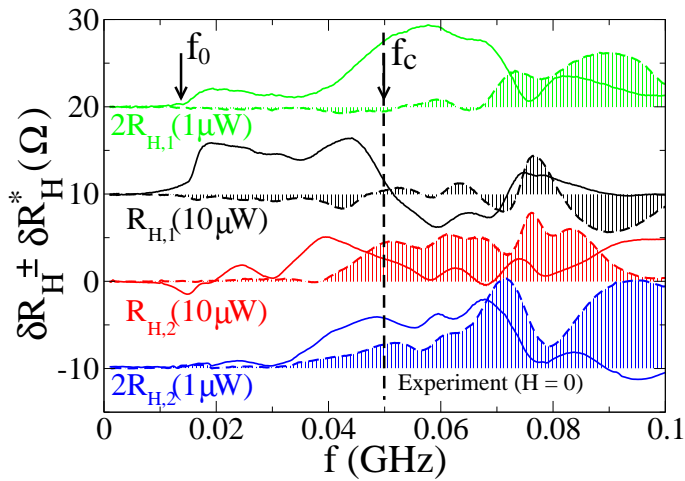}
\leavevmode
\caption{ (Color online) Change of Hall resistance on sample $(A)$ 
under microwave irradiation on the local split-gate $S$ as a function of microwave frequency. 
Continuous curves represent the Onsager-symmetric part of the Hall resistance $\delta R_{H} + \delta R^*_{H}$, while 
the dashed curves show the anti-symmetric part 
$\delta R_{H} - \delta R^*_{H}$. The data was obtained on the two Hall probes $\delta R_{H,1}$ and $\delta R_{H,2}$ 
at injected microwave powers of $1 {\rm \mu W}$ and $10 {\rm \mu W}$.
The origin of the different curves is shifted by an arbitrary offset for clarity ($\delta R_{H}$ vanishes at low frequencies $f < 1$ MHz) 
and the values of $\delta R_{H}$ at power $1 {\rm \mu W}$ are scaled upwards by a factor $2$.
\label{peaksg}}
\end{center}
\end{figure}

Our theoretical model predicts that the onset of time-reversal symmetry breaking by an
ac-radiation can be probed directly by measurements of the difference $R_H - R_H^*$ as a 
function of ac-frequency without introducing an external magnetic field. 
We have checked this prediction experimentally by applying a high frequency 
potential on the split gate $[S]$ on sample $(A)$. 
In order to remove the contribution of geometrical imperfections of our Hall junctions
that lead to non-zero $R_H$ even in the absence of magnetic field we now focus 
on the difference $\delta R_H$ between the Hall resistance with AC-driving and its equilibrium value. 
We have measured the change of Hall resistances for the two Hall junctions of the sample, polarized in the two reciprocal configurations: 
$\delta R_{H,1}, \delta R_{H,1}^*, \delta R_{H,2}$ and $\delta R_{H,2}^*$
as a function of microwave frequency $f$ for fixed injected microwave power. 
The data, shown on Fig.~3, indicate the following scenario. At very low driving 
frequencies $f < f_0 \simeq 10$MHz both symmetric and anti-symmetric components 
$\delta R_{H,i} + \delta R_{H,i}^*$ and $\delta R_{H,i} - \delta R_{H,i}^*$ are zero ($i = 1,2$), 
we attribute this to the fact that our capacitive coupling is not efficient at so low frequencies 
and the amplitude of the AC potential is very small in this limit. 
For higher frequencies, a change of Hall resistance due to microwave irradiation is observed, however 
as expected from our model the reciprocity relations is still valid $\delta R_{H,i} \simeq \delta R_{H,i}^*$.
It is only for $f > f_c \simeq 50\; {\rm MHz}$ that the anti-symmetric component becomes significant, 
and for higher frequencies (we measured up to $f = 10\;{\rm GHz}$) we observe that the 
symmetric and anti-symmetric components are of the same order of magnitude. We find that 
that the critical frequency $f_c$ is similar for both Hall junctions and weakly depends on injected microwave power. 
This is consistent with our simulations where the threshold $\omega_c$ did not depend on the potential amplitude $U_0$. 
We note that as in our theoretical results on Fig.~2, $\delta R_H + \delta R_H^*$ scales proportionally to power ($U_0^2$) 
for $f < f_c$. At higher higher frequencies a more complicated behavior is observed since $\delta R_H + \delta R_H^*$
may change sign as a function of frequency.
We showed that in a ballistic sample the length $L_c = \frac{v_F}{\omega_c}$ is of the order 
of the distance between reservoirs $L$, 
in a diffusive sample with mean free path $l_e$ (we estimate $L \simeq 65\;{\rm \mu m}$ and $l_e \simeq 10\;{\rm \mu m}$ for our samples)
we expect that the relevant length scale is determined by $L_c \simeq L^2 / l_e \simeq 400\;{\rm \mu m}$.
Such a value for $L_c$ is consistent with a critical frequency $f_c = 50\; {\rm MHz}$ as observed in the experiment. 
We remark that the onset of the difference $\delta R_H - \delta R_H^*$ could also be caused by 
the appearance of stationary orbital magnetism under microwave irradiation, an effect that was predicted theoretically in Ref.~\cite{toulouse}.
Indeed the quantity $R_H - R_H^*$ is proportional to the induced magnetic field. 
However the magnetic field required to change the value of Hall resistance by $\delta R_H \simeq 10\;{\rm \Omega}$ as observed on Fig.~3 
is $H \simeq 30\;{\rm G}$. This is several orders of magnitudes larger than the effect predicted in \cite{toulouse},
which under our experimental conditions should create magnetic fields of the order of $H \simeq 10^{-2}\;{\rm G}$. 

\begin{figure}
\begin{center}
\includegraphics[clip=true,width=9cm]{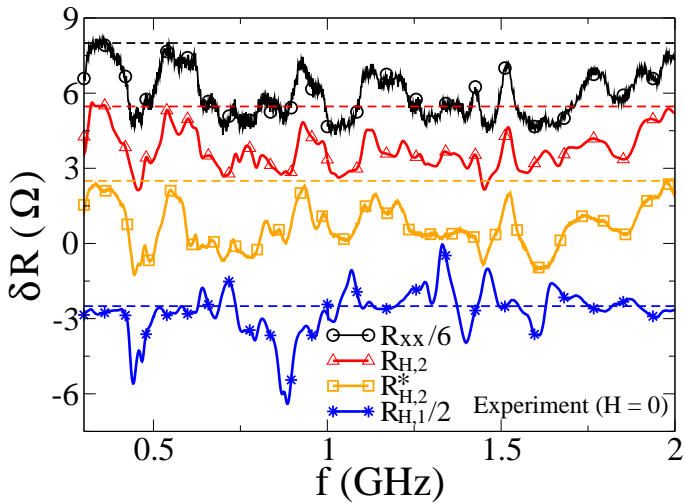}
%\leavevmode
%\epsfxsize=8cm
\caption{(Color online) Change of Hall and longitudinal resistance under homogeneous microwave irradiation of sample $(B)$ 
as a function of microwave frequency at injected microwave power of $10\;{\mu W}$ and $T = 0.3\;{\rm K}$. 
The curves are marked with circles for $\delta R_{xx} / 6$, triangles for $\delta R_{H,2}$, squares for $\delta R^{*}_{H,2}$ 
and stars for $\delta R_{H,1}/2$. The origin ($\delta R = 0$) is indicated by the dashed lines and is shifted for clarity.
Note that the signal measured on $R_{H,2}$ closely follows $R_{xx}$.} 
\label{scalingdynamicCB}
\end{center}
\end{figure}

We now show that the spatial inhomogeneity of the alternating electric  field is essential in order to observe
strong deviations from Onsager symmetries.  We prove this experimentally by irradiating  
sample $(B)$ with an  external electromagnetic field homogeneous on the sample scale emitted 
with a macroscopic antenna. 
For the Hall junction $R_{H,1}$ the presence of a quantum dot a few microns away from the sample (see Fig.~1)
is expected to deform the external potential creating inhomogeneities in the electric field.
On the contrary for $R_{H,2}$ we expect an homogeneous irradiation. 
On Fig.~4, we compare the variation of Hall resistances $\delta R_{H,2}, \delta R_{H,2}^*$ with 
frequency $f$ at fixed power.
We find that the Onsager relations are verified even in the limit of very high frequencies $f \simeq  1\;{\rm GHz}$ compared to our 
previous experiment, and the relation  $\delta R_{H,2} \simeq \delta R_{H,2}^*$ is valid. 
We also note that $\delta R_{H,2}$ is proportional 
to the change of sample resistance, $\delta R_{xx}$ which is negative at all frequencies. 
The sign of $\delta R_{xx}$ corresponds 
to  heating, since we are in a regime where the sample resistivity decreases with temperature ($T \simeq 0.3\;{\rm K}$). 
The proportionality between $\delta R_{xx}$ and $\delta R_{H,2}$ can be explained as a 
geometrical offset in $R_{H,2}$ proportional to $R_{xx}$. In fact, such a simple proportionality 
relation is a good indication of the external electric field homogeneity in the Hall junction.
It does not hold for inhomogeneous irradiation, as shown by our measurements on $R_{H,1}$.
In this case the proportionality to $\delta R_{xx}$ is not observed
and we find  $\delta R_{H,1} \ne \delta R^*_{H,1}$ (data not shown) as in sample $(A)$. 

In conclusion we have addressed the validity of reciprocity relations in a Hall bar under AC driving. 
We have established that the magnetotransport in our samples is well 
described by the billiard model of \cite{Beenakker}. We have generalized this model to include the effect of an AC field.
With this model we predicted the onset of deviations from reciprocity 
relations at high enough AC frequencies even at zero magnetic field. 
We have checked this prediction experimentally by applying an inhomogeneous AC-field on the Hall bar. 
The transition from the low frequency 
regime where reciprocity symmetry holds to the asymmetric regime at high frequencies was 
clearly observed. Finally by irradiating a Hall bar with a macroscopic antenna, 
we established that the reciprocal relations are more robust 
under homogeneous irradiation.
Intriguingly the signal we observe: $R_H - R_H^*$ resembles a static magnetic field except 
for its very large amplitude. This raises the question of a detector that can discriminate 
between inhomogeneous AC-electric fields and a small static magnetic field. 

We are grateful to A. Cavanna,  B. Etienne and U. Gennser for the $GaAl/Ga_{1-x}Al_{x}As$ heterojunction and to 
S. Gu\'eron, J. Gabelli, F. Pierre, M. Monteverde, C. Ojeda, R. Deblock, V. Kravtsov, M. Polianski for fruitful discussions.
We acknowledge ANR NANOTERA and DGA for support.

\end{document}